# Moderate-Resolution Holographic Spectrograph

E. R. Muslimov,[1] N. K. Pavlycheva,[1] G. G. Valyavin,[2] and S. N. Fabrika[2, 3]
[1]Kazan National Research Technical University named after A.N.Tupolev (KAI), Kazan, 420111 Russia
[2]Special Astrophysical Observatory of the Russian AS, Nizhnij Arkhyz 369167, Russia
[3]Kazan (Volga region) Federal University, Kazan, 420008 Russia

**Abstract**. We present a new scheme of a moderate-resolution spectrograph based on a cascade of serial holographic gratings each of which produces an individual spectrum with a resolution of about 6000 and a bandwidth of 80 nm. The gratings ensure centering of each part of the spectrum they produce so as to provide uniform coverage of the broadest possible wavelength interval. In this study we manage to simultaneously cover the 430–680 nm interval without gaps using three gratings. Efficiency of the spectrograph optical system itself from the entrance slit to the CCD detector is typically of about 60% with a maximum of 75%. We discuss the advantages and drawbacks of the new spectrograph scheme as well as the astrophysical tasks for which the instrument can be used.

## 1 Introduction

This document shows the suggested format and appearance of a manuscript prepared for SPIE journals. Accepted papers will be professionally typeset. This template is intended to be a tool to improve manuscript clarity for the reviewers. The final layout of the typeset paper will not match this template layout. Spectroscopy of faint objects on major optical telescopes is performed using low- and intermediate-resolution spectrographs with the resolution $R=\lambda/\Delta\lambda$ ranging from one hundred to two-three thousand. As efficiency standards we can use the most well-known instruments of this class – FORS spectrographs on VLT telescopes of European Southern Observatory [1,2] and FOCAS spectrograph of Subaru telescope of NAOJ observatory [3]. A similar spectrograph – SCORPIO [4] – has comparable parameters and is used extensively on the 6-m telescope of the Special Astrophysical Observatory of the Russian Academy of Sciences. These spectrographs demonstrate close-to-limiting efficiency. In particular, FOCAS spectrograph has an optical throughput from the entrance slit to the CCD detector (CCD excluded) of up to 82\% with a full coverage of the entire optical wavelength range. This result leads us to conclude that spectrographs operating in the mode of low spectral resolution, on the whole, require no conceptual changes.

For observing bright objects, higher-resolution ($R > 10\,000$) spectrographs are used, which are based on more sophisticated schemes involving echelle gratings. The main advantage of an echelle spectrograph is its high spectral resolution combined with relatively small diameter of the collimated beam [5]. However, similar schemes have limited throughput of the optical train. The throughput is typically equal to only about 8% [6] and can amount to 15-20% [7,8] for modern instruments, which is considered as the technological limit. The wide gap between the quantum efficiency of low/intermediate and high-resolution spectrographs pushes one to consider alternative solutions, which could provide complete coverage of the optical wavelength range in the mode of intermediate or higher spectral resolution. In this paper we present the scheme of one of such possible solutions based on volume phase diffraction gratings.

It is well known that a thick volume phase diffraction grating may have high (up to 100%) diffraction efficiency in a narrow spectral interval. Such a grating can simultaneously serve as a dispersion element and a spectral filter [9]. These properties of volume phase gratings can be used to construct a scheme of a spectrograph with a very high throughput. Some of the potentialities and advantages of the use of such optical elements in astronomical spectrographs were demonstrated, e.g., by [10,11].

The proposed optical scheme of the spectrograph is based on the use of a cascade of several volume phase diffraction gratings. Each grating of the cascade produces a spectrum in the individual spectral range similarly to how a diffraction order of the echelle grating produces its own part of the image of the spectrum. A cascade of gratings mounted in series ensures the centering of each spectral interval produced so as to provide uniform coverage of the broadest possible spectral range. We show that such a concept allows the spectral resolution to be increased severalfold with no substantial loss of efficiency while maintaining the registered spectral range typical for intermediate- and low-resolution spectroscopy. We also demonstrate, relying on energy estimates of the efficiency of the optical scheme, that the presented solution is an efficient and inexpensive alternative for intermediate-resolution spectrographs with spectral resolutions $R$ ranging from 5000 to 10 000. The proposed system has a substantially higher efficiency (up to 75% at the maxima of grating orders) compared to intermediate-resolution echelle spectrograph.

The instrument is currently developed within the framework of a program involving the following research directions: search for very massive stars and intermediate-mass black holes, study of exoplanets, and magnetometry of stars. These directions (especially the first two of them) require an efficient intermediate-resolution spectrograph with maximum throughput of the optical train. It is also evident that an efficient intermediate-resolution spectrograph will be needed not only for research in the above fields, but also for many currently important fields of modern astrophysics. This study is limited to presenting the concept of the instrument without considering any particular scheme of the telescope that we plan to use in combination with the spectrograph. In Section 1 we present the optical scheme of the spectrograph and discuss its details. In Section 2 we discuss the diffraction efficiency of selective holographic gratings as a function of wavelength in the context of their use in astronomical spectroscopy. In Sections 3 and 4 we list the topical problems for intermediate-resolution spectroscopy, present the main conclusions of this study, and discuss the advantages and drawbacks of the system and further steps towards its improvement.

## 2 Optical scheme of the spectrograph

In the proposed scheme several gratings are located in series in the collimated beam and each grating operates in a certain wavelength interval. Each grating is a glued unit made up of two wedges. The grating is applied to the inner surface of the unit with a tilt in the meridional and sagittal planes. The outer edges of the wedges make up a plane-parallel plate. In the +1$^{st}$ order of diffraction such a grating produces the spectrum in the dedicated spectral range, in the 0$^{th}$ order of diffraction it works as a plane-parallel plate and has little effect on the propagation of other spectral components of radiation. The tilt of the grating surface in the meridional plane allows matching the centers of the images of the spectra for different bands taking into account their different dispersions, and the tilt in the sagittal plane makes it possible to produce several rows in the image of the spectrum corresponding to different spectral bands. The preferred choice for the collimating and camera objectives are achromatized lens systems with the highest possible throughput. To illustrate the proposed design principle of the optical scheme we show in Fig.1 the white-light source photos obtained in the 0$^{th}$ and +1$^{st}$ orders using a typical volume holographic grating.

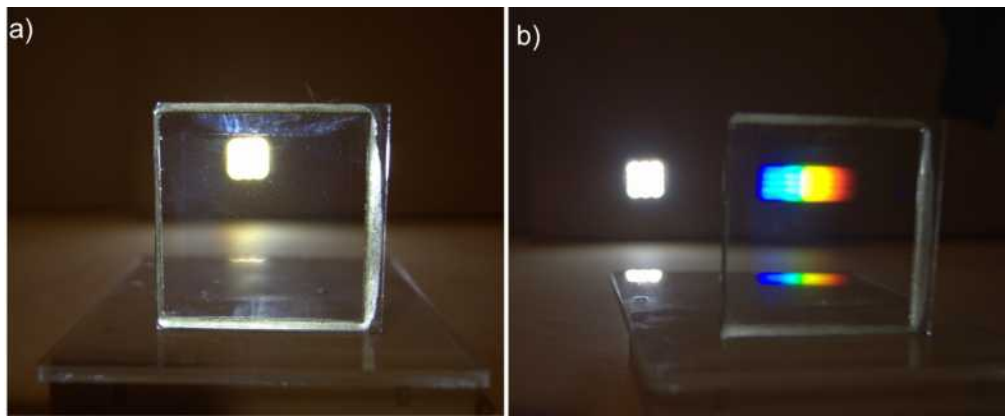

**Fig. 1** Photo of a LED cluster taken through a typical volume phase holographic grating in the 0$^{th}$ (a) and +1$^{st}$ (b) orders of diffraction. The observed images demonstrate dispersing and spectrum-splitting properties of volume holographic grating.

The facility uses a 40x40 mm grating with a groove density of 1200 grooves·mm$^{-1}$ recorded on layers of dichromated gelatin (DCG) using standard technology; the light source consists of a cluster of 12 LEDs installed parallel to the grating surface. The photos clearly show that in the $0^{th}$ order of diffraction observed along the normal to the grating surface (Fig.1a), non-dispersed radiation propagates without distortions, whereas in the $+1^{st}$ order observed at an angle of about 41.3° (Fig.1b), the spectrum of the source forms. Thus each volume holographic grating works simultaneously as a dispersing and spectral-splitting element.

We use the following algorithm to compute the optical scheme of the spectrograph. The initial data consist of: spectral operating range, length of the scheme, size of the detector and reciprocal linear dispersion. At the first stage, the spectral bands are determined into which the entire working wavelength range is subdivided, and the focal distance of objective lenses is chosen. Then for each grating the groove density and turn angle in two planes are computed that ensure the required dispersion and row separation. At the next stage the design parameters of the objectives are determined and aberration computations are performed for them. Computation practice shows that the parameters of the objective lens systems should be further optimized for the assembled scheme. The image quality parameters are computed for the defined parameters. At the final stage of the computations the parameters of the holographic layer of each grating are optimized for operation in the chosen spectral band. The concept considered here assumes that diffraction on grating is practically unobserved outside the selected band. It makes sense to perform control simulation using some numerical method to verify the fulfillment of this condition.

Let us now consider the optical scheme of a spectrograph for the visual range, 430-680 nm. We subdivide the spectral interval into equal spectral bands 430-513, 513-597, and 597-680 nm. Our design computations are based on the parameters of commercially manufactured CCD array detectors with a 36x24 mm sensitive chip and a pixel size of 5.2x5.2 μm (note that in the final version of the instrument the actual format will be chosen so as to match the standards adopted at the observatories where the spectrograph can be used). Let us set the target length of the spectrum image equal to 30~mm for each band. Based on the size limits and requirements to linear dispersion we set the focal distances of the collimating and camera objectives equal to 170 mm. We chose the relative aperture of the objectives based on the condition of aperture matching with the optical system of the 6-m telescope of the Special Astrophysical Observatory of the Russian Academy of Sciences in the primary focus and with the slit-width ratio equal to one. We adopt the numerical aperture equal to 0.13, which is even somewhat greater than that of the 6-m Telescope [12]. Here it should be pointed out that implementing the optical scheme of a spectrograph with the chosen parameters of the incoming beam is a rather difficult task. We chose such a combination of parameters to demonstrate the practical feasibility of the instrument even in the case where it is recommended for mounting in the foci of telescopes with relatively large numerical apertures and within tight space (e.g., in the primary focus of the 6-m telescope). However, after the decision is made about the particular telescope and focus where the spectrograph is to be installed the parameters of aperture matching will be adjusted depending in the particular scheme.

Groove density and its tilt angle in the meridional plane can be approximately determined from the following set of equations:

$$\begin{aligned}
\phi'_{12} &= \phi'_{22}, \\
\phi'_{12} &= \phi'_{32}, \\
f'(\tan(\phi'_{13} - \phi'_{12}) - \text{tg}(\phi'_{11} - \phi'_{12})) &= \\
f'(\tan(\phi'_{23} - \phi'_{22}) - \text{tg}(\phi'_{21} - \phi'_{22})), \\
f'(\tan(\phi'_{13} - \phi'_{12}) - \text{tg}(\phi'_{11} - \phi'_{12})) &= \\
f'(\tan(\phi'_{33} - \phi'_{32}) - \text{tg}(\phi'_{31} - \phi'_{32})), \\
f'(\tan(\phi'_{13} - \phi'_{12}) - \text{tg}(\phi'_{11} - \phi'_{12})) &= L_s.
\end{aligned} \quad (1)$$

where $\varphi'_{ij}$ is the diffraction angle for wavelength $j$ of spectral band $i$, which is equal to

$$\phi'_{ij} = \arcsin(\lambda_{ij} N_i + \sin \phi_i) \tag{2}$$

Here $N_i$ and $\varphi_i$ are the groove density and incidence angle of the corresponding grating, respectively; $f'$, the focal distance of the camera objective, and $L_s$, the length of the row in the image of the spectrum.

We set a small tilt angle of the first grating, 10°, and determine the groove density values for the initial approximation: $N_1$=1405.4 lines·mm$^{-1}$, $\varphi_1$=-6.694°; $N_2$=1386.7 lines·mm$^{-1}$, $\varphi_2$=-0.591°; $N_3$=1401.4 lines·mm$^{-1}$, $\varphi_3$=6.554°. We assume that the distance between the rows in the spectrum image should be no less than 2.2 mm, and perform numerical optimization of the system to determine the tilt angles of the gratings in the sagittal plane and the final groove density and turn angles in the meridional plane: $N_1$=1616 lines·mm$^{-1}$, $\varphi_1$=-10.007°, $\gamma_1$=13.951°; $N_2$=1337 lines·mm$^{-1}$, $\varphi_2$=1.258°, $\gamma_2$=9.116°; $N_3$=1049 lines·mm$^{-1}$, $\varphi_3$=15.053°, $\gamma_3$=14.612°.

The camera and collimating objectives are made in accordance with the classical triplet scheme. Both objectives are optimized taking into account the specific conditions of the operation: the collimating objective to provide extended spectral range and narrow field of view, and the camera objective, to provide forward-offset entrance pupil and wide field of view. To correct the residual aberrations of the camera objective an extra lens has to be introduced into its scheme.

Figure 2 shows the general view of the spectrograph scheme after optimization. The diffraction gratings in the beam are mounted in reverse order - the first is the grating for the long-wavelength band 597-680 nm, and the last one, for the short-wavelength band 430--513~nm. The overall size of the system is 755x440x170 mm. To assess the image quality achievable in this scheme, let us consider the spot diagrams of the spectrometer (Fig.3).

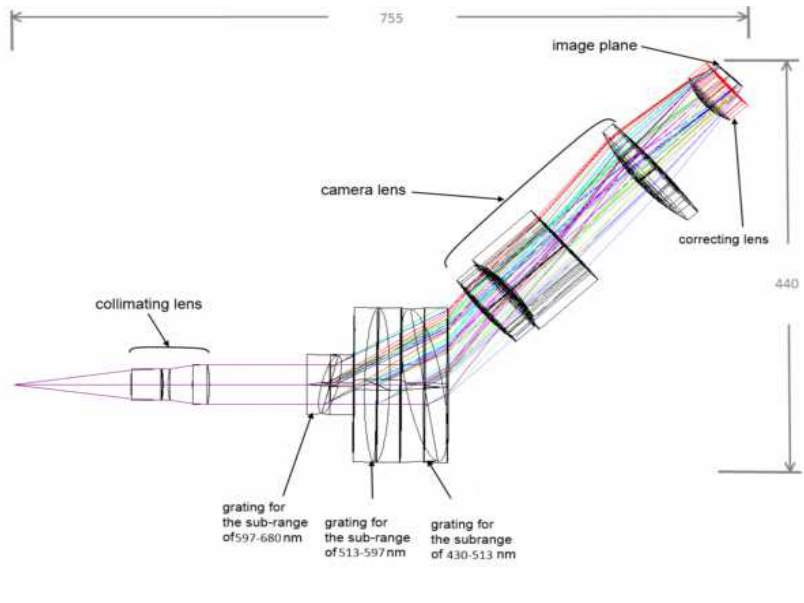

**Fig. 2** Principal optical scheme of the spectrograph. Each of the three volume phase diffraction gratings of the cascade is placed between two identical wedges. Because of the high spectral selectivity of the volume structure such a grating produces the spectrum in the selected wavelength band (430-513, 513-597, and 597-680 nm), and works as a plane-parallel plate outside this band. The numerical aperture of the instrument is equal to 0.13, and the focal lengths of the camera and collimating objectives are equal to 170 mm.

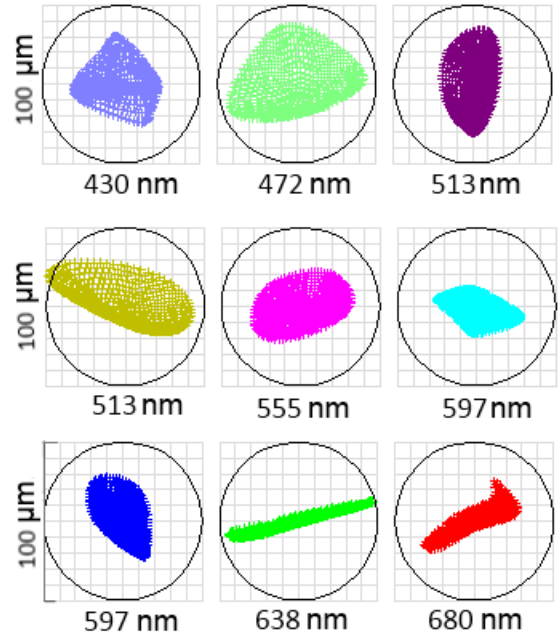

**Fig. 3** Scatter diagrams of the spectrograph for the edge and middle working wavelengths. The sizes of the scatter diagrams are indicative of a rather high resolution limit, which varies only slightly along the spectrum.

Figure 4 shows the general view of the spectrum image produced by the spectrograph. The reciprocal linear dispersion is equal to 2.61 nm·mm$^{-1}$ for the short-wavelength band (430-513 nm), 2.76 nm·mm$^{-1}$ for the mid-wavelength band (513-597 nm), and 2.86 nm·mm$^{-1}$ for the long-wavelength (597-680nm) bands.

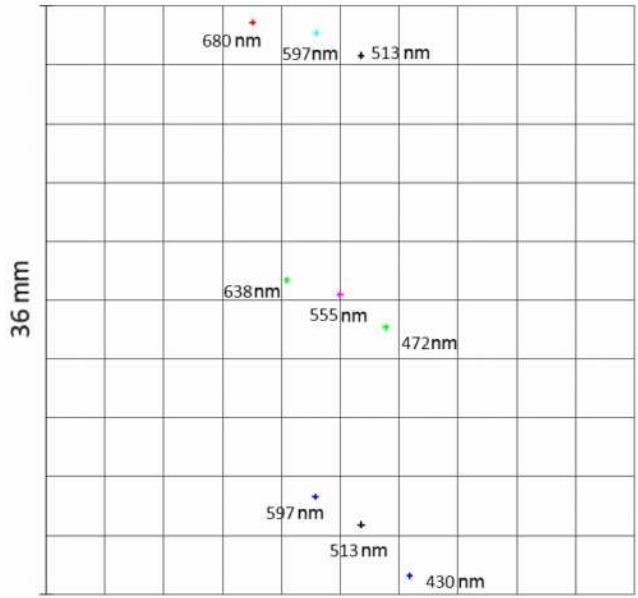

**Fig. 4** Filling scheme of the field of view of the spectrograph. The scheme shows monochromatic images of the entrance slit for the minimum-, mid-, maximum-wavelength operating bands: 430-513, 51--597, and 597-680 nm, respectively. Centering of the rows of the spectral ecomposition (arranged vertically in the scheme) is provided by the turn of the surfaces of each diffraction grating in the meridional plane. The distance between the decomposition rows is no less than 2.2 mm and is ensured by the turn of the gratings in the sagittal plane. The frame size is 36x24 mm.

To determine the spectral resolution provided by the given scheme, let us determine the instrumental functions (IF) of the spectrograph. The instrumental functions are computed for a 30 μm wide entrance slit. Note that to ensure matching of the entrance slit of the spectrograph with the image produced by a particular telescope (e.g., the 6-m telescope), it may be necessary to use an image slicer similar to that described by [13]. Note again, however, that implementing the instrument on a particular telescope with particular foci and allowed collimated beam diameters requires a revision of the beam transformation parameters adopted here in accordance with the size of the entrance slit and the need to use an image slicer. We defer this problem to future studies. Figure 5 shows the plots of the instrumental functions for the adopted wavelengths. The spectral resolution determined from the width of the IF at half maximum given the reciprocal linear dispersion values is equal to: 0.082-0.147,0.083-0.098, and 0.086-0.124nm for the short-, mid-, and long-wavelength bands, respectively. Hence the maximum resolution [14] of the spectrograph for each wavelength band is equal to 5243, 7192, and 7906, respectively. It can be easily seen that these values are appreciably lower than the limiting resolution for the given dispersion and collimated beam diameter. Hence the resolution of the spectrograph is determined primarily by the finite width of the entrance slit and aberrations of the objectives. We now have to determine the optimum parameters of the holographic layer for each grating.

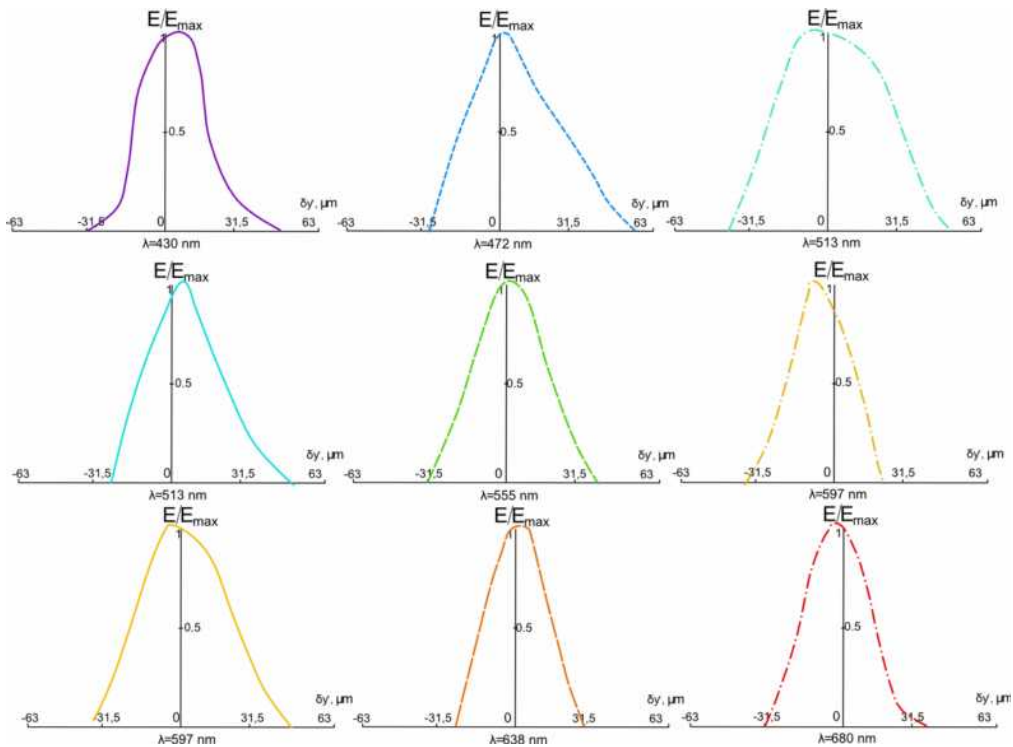

**Fig. 5** Instrumental functions of the spectrograph for the edge and mid wavelengths of the operating spectral bands of the instrument. The width of the entrance slit is 30 μm. The linear resolution limit defined as the FWHM of the instrumental function is equal to 31-56 μm, 31-45 μm, and 30-34 μm for the short-, mid-, and long-wavelength bands, respectively. The corresponding spectral resolution is equal to 0.082-0.147, 0.083-0.098, and 0.086-0.124 nm, respectively.

# 3 Diffraction efficiency of the volume-phase gratings

To compute the parameters of the holographic layer of volume phase gratings (thickness $t$ refraction index modulation depth $\Delta n$), we use the coupled wave theory developed by Kogelnik [15]. This theory is based on a number of assumptions, however, it can be used to compute the diffraction efficiency of a volume hologram using simple analytical relations.

We assume that each grating can be recorded by two collimated beams. The position of the grating fringes then depends on the incidence angles of the recording beams:

$$\psi = 90° - \frac{\arcsin(\sin i_1/n) - \arcsin(\sin i_2/n)}{2}, \qquad (3)$$

where $i_1$ and $i_2$ are the incidence angles and $n$, the refractive index of the holographic layer (for the most widespread material - dichromated gelatin - we can adopt $n=1.51$). The tilt angle of the fringes should conform the Bragg condition

$$2\cos(\psi - \phi) = \frac{N\lambda_2}{n}, \qquad (4)$$

Here $\lambda_2$ is the center wavelength of the corresponding spectral band. We should also take into account the relation between the incidence angles of the recording beams and the grooves density of the grating:

$$\sin i_1 - \sin i_2 = N\lambda_0, \qquad (5)$$

where $\lambda_0$ is the recording wavelength. Having determined the incidence angles of the recording beams and the tilt of the grating fringes from equations (4)-(6}, we can pass to optimizing the parameters of the layer. To this end we compose the merit function of the form:

$$f_m(t, \Delta n) = \sum_{p=1}^{k} \left(\eta_s(t, \Delta n, \lambda_p) - \eta_{\text{tar}}(\lambda_p)\right)^2, \qquad (6)$$

where $\eta_s$ is the diffraction efficiency for unpolarized radiation determined by the formulas of Kogelnik's theory and $\eta_{tar}$ is the target value of the diffraction efficiency determined by rectangle function

$$\eta_{\text{tar},i}(\lambda) = \begin{vmatrix} 1, \lambda_{i1} \leq \lambda \leq \lambda_{i3}, \\ 0, \lambda < \lambda_{i1} \cup \lambda > \lambda_{i3}. \end{vmatrix} \quad (7)$$

We then minimize function (6) in the domain of technologically feasible values of the layer thickness and modulation depth, 10μm<t<10μm, 0.0005<Δn<0.13 [11,16], to determine the optimum values of the above parameters for each grating: t=11μm, Δn=0.021 for the short-wavelength band t=15μm, Δn=0.016, for the mid-wavelength band, and t=28μm, Δn=0.011, for the long-wavelength band. Figure 6 shows the resulting wavelength dependences of the diffraction efficiency.

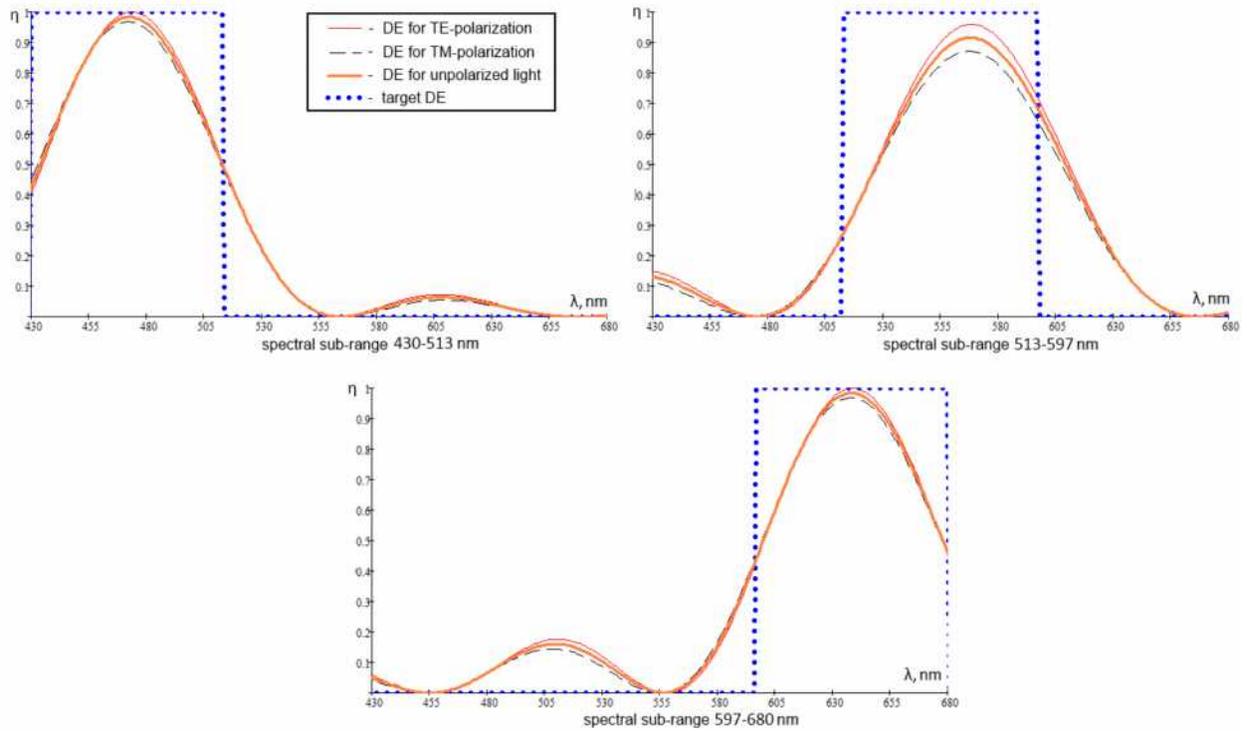

**Fig. 6** Spectral dependences of the diffraction efficiency of volume phase gratings of the spectrograph. The dependence of diffraction efficiency was computed and optimized using the analytical relations of Kogelnik's coupled wave theory [10].
13

The results reported above were obtained without taking into account the turn of the gratings in the sagittal plane. In reality the turn angles are finite and to determine the diffraction efficiency we have to numerically model the case of conical diffraction. For the parameters computed above such a modeling was performed using RCWA (rigorous coupled wave analysis) method [17]. The results of modeling qualitatively confirm the results of the analytical calculation - the gratings have high diffraction efficiency in a relatively narrow spectral band, which is close to the computed band. However, because of the turn in the sagittal plane and influence of nonworking orders of diffraction the operating spectral band of the gratings shifts toward shorter wavelengths and the maximum diffraction efficiency decreases. Figure 7 shows, by way of an example, the plots of the spectral dependence of the diffraction efficiency for the mid-wavelength band obtained using numerical simulations and analytical computations.

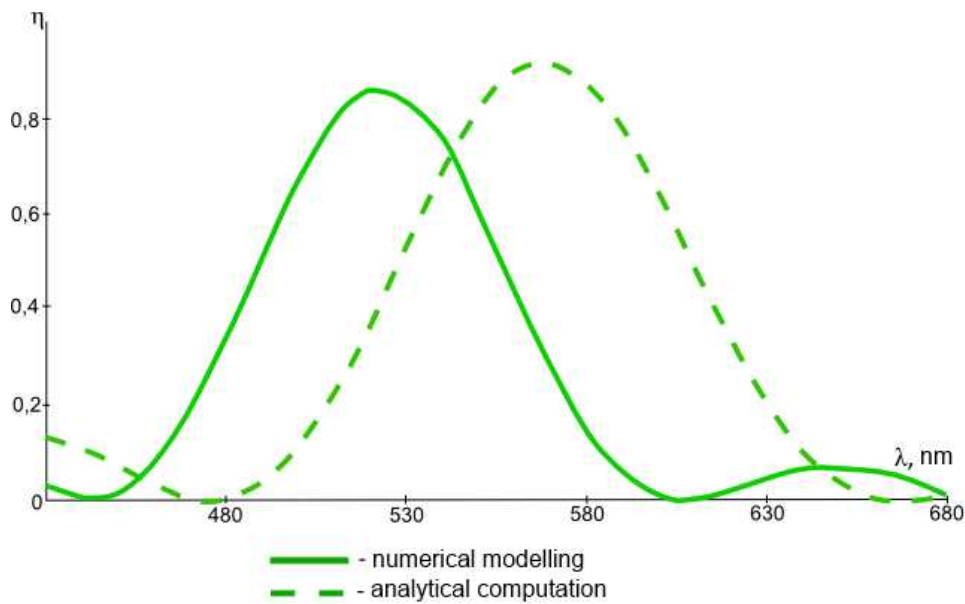

**Fig. 7** Wavelength dependence of the diffraction efficiency of the grating of the mid-wavelength channel of the spectrograph. The solid line corresponds to the dependence computed using the method of rigorous coupled wave analysis and the dashed line, to the calculation based on analytical formulas. The results of the computations agree qualitatively in the context of the formulated roblem. The shift of the position of the maximum and its magnitude (~5%) can be explained by the effect of conical diffraction.

Thus the results of simulations confirm that it is possible to make a spectrograph with selective volume phase gratings. In this case to accurately determine the parameters of the holographic layer of each grating, optimization should be performed based on a precise numerical model.

## 4 Priority astrophysical problems for moderate-resolution spectroscopy

In this part we briefly outline the most topical astrophysical problems that can be solved using spectrographs of the type described above. These problems include the study of the nearby Universe galaxies, and stars in these galaxies. At such distances massive stars can be seen as individual objects. The chemical evolution of galaxies is fully determined by the star-formation rate, or, more precisely, by the number of massive stars in the galaxy. The evolution of massive stars is, in turn, determined by their chemical composition, which determines the mass-loss rate in their winds [18]. This process is a factor on which the black holes and neutron stars formation rate and numbers in galaxies depend. Of fundamental importance for astrophysics is to know which relativistic objects represent the last stage in the life of stars of particular type: black hole, neutron star, or the accompanying explosion of the star leaves no remnant at all (``pair instability SN'' [19]). It is important to perform observations at all stages of the evolution of massive stars - from the main sequence to the LBV (``Luminous Blue Variables'') stage, WR (Wolf-Rayet), supernova explosion, and, finally, to the formation of x-ray sources (black holes and neutron stars in binary systems). Optical ``counterparts'' of x-ray sources (usually 17-13$^m$ objects) are quite accessible for moderate-resolution spectroscopic observations. Transition from low($R$~1000) to moderate ($R$~5000-10 000) resolution in the spectroscopy of faint objects is of fundamental importance, because it will allow resolving lines in the spectra of massive stars, lines in the spectra of galaxies, and active galactic nuclei (AGN).

The proposed spectrograph can be especially efficient when used for the study of supernovae, optical counterparts of gamma ray bursts (it is important to obtain the spectra of such objects within several hours after they appear in the sky), central regions of young star clusters, ultra- and hyperbright x-ray sources. A spectral resolution equivalent to 50 km s$^{-1}$, allows analyzing the profiles of emission and absorption lines of the winds of massive LBV and WB stars and emission supergiants. Also of current importance is the search for very massive stars (VMS). So far stars with masses up to 200 solar masses have been discovered [20]. Such stars can form in centers of young star clusters. Their evolutionary lifetime is shorter than three million years. The star can then become a so called intermediate-mass black hole (IMBH). Establishing the very fact that IMBHs exist would be an important contribution to various fields of modern astrophysics: formation of the first stars during the first 500 million years after the Big Bang, formation of dwarf galaxies, star clusters, and VMS.

Spectrographs of this type operating in a broad wavelength range can be used to study low-mass compact objects in binary systems (cataclysmic variables) and single objects like white dwarfs. Studies of magnetically induced spot irregularities at the surfaces of white dwarfs [21] and their chemical composition and masses require an efficient spectrograph. Another important task is the search for and the study of exoplanets. Such a spectrograph, as expected, will have an optical throughput of no less than 50%, and will make it possible to obtain spectra of exoplanets based on the signal reflected from their surfaces in the spectral lines of their parent stars [22,23] even when operated on small telescopes with a 2-m primary mirror.

## 5 Discussion and conclusions

In this paper we propose a concept of a spectrograph with high optical throughput and moderate spectral resolution to be used for astrophysical studies. The optical scheme of the spectrograph is based on volume phase holographic diffraction gratings, which work as dispersing elements and spectral filters. We report the results of the computation and analysis of the scheme of such a spectrograph for the 430-680nm wavelength interval. The spectrograph produces the image of the spectrum in the form of three rows corresponding to the 430-513, 513-597, and 597-680nm wavelength bands, and has a resolution of 5200 to 7900 depending on the band. The results of the computation of the diffraction efficiency of the gratings show that high efficiency in the chosen band and high spectral selectivity of grating can be achieved with technologically feasible parameters. The accurate determination of the parameters of the layer requires optimization using precise numerical methods of diffraction modeling. The advantages of the scheme developed include the rather high spectral resolution, high transmission of the optical train, moderate size, and the use of rather unsophisticated optical components. The maximum throughput of the optical train at the centers of optical bands reaches 75% with the nominal efficiency equal to 40-50%.

The shortcomings of the new scheme are the fast increase of the clear apertures of the components with increasing dispersion and/or increasing number of lines of the decomposition, which limits further increase of the spectral resolution. Another important drawback of the scheme is the diffraction outside the working band, which can result in appreciable losses in the illumination of the spectrum (primarily at the edges of the decomposition lines) and stray illumination of the detector. However, the problem of the control of the diffraction efficiency profile of volume phase holographic optical elements was earlier solved successfully in terms of various applications. In particular, [24] describes a technique of the computation and manufacturing of volume phase holographic filters with a narrow operating spectral band, sharp boundaries, and variable width. [25] reports the results of the computation of the diffraction efficiency of a reflective volume phase grating, and demonstrate its high spectral selectivity and steepness of the response curve at the boundaries of the operating wavelength band. A certain drawback of the reported scheme is the small width of the slit, which ensures the required spectral resolution. It can result in the need to introduce an image slicer if the instrument is supposed to be operated in the present configuration on a large telescope. However, as we pointed out above, this problem can be solved by revising the current parameters of the scheme. Furthermore, the image slicer manufacturing technologies are well developed and can also be used.

Note also that at this stage of the work we purposefully did not consider the technologies of the production of volume phase gratings. We assumed that the gratings should be recorded in parallel beams and should therefore have a sine distribution of the refraction index. On the other hand, there are dedicated schemes, which allow the modulation of the refraction index and the diffraction efficiency profiles to be controlled. Examples of such schemes include recording in opposite beams propose in [26] and multiexposure record described in [27].

The proposed scheme shows that the drawbacks of the system revealed will be successfully eliminated as the project progresses.

*Acknowledgments*

We thank J.-P. Hugonin (Institut d'Optique, Palaiseau, France) for providing Reticolo software, which allows modeling diffraction on arbitrary periodic structures. This research was supported by the Russian Science Foundation (project No.14-50-00043).